# Linear magnetoresistance in the low-field limit in density-wave materials

Yejun Feng[1], Yishu Wang[2], D.M. Silevitch[2], J.-Q. Yan[3], Riki Kobayashi[1,4], Masato Hedo[4], Takao Nakama[4], Yoshichika Ōnuki[4], A. V. Suslov[5], B. Mihaila[6], P. B. Littlewood[7], T. F. Rosenbaum[2]

[1]Okinawa Institute of Science and Technology Graduate University, Onna, Okinawa 904-0495, Japan

[2]Division of Physics, Mathematics, and Astronomy, California Institute of Technology, Pasadena, CA 91125, USA

[3]Materials Science and Technology Division, Oak Ridge National Laboratory, Oak Ridge, TN 37831, USA

[4]Faculty of Science, University of the Ryukyus, Nishihara, Okinawa 903-0213, Japan

[5]National High Magnetic Field Laboratory, Tallahassee, FL 32310, USA

[6]National Science Foundation, Arlington, VA 22230, USA

[7]The James Franck Institute and Department of Physics, The University of Chicago, Chicago, IL 60637, USA

**The magnetoresistance (MR) of a material is typically insensitive to reversing the applied field direction and varies quadratically with magnetic field in the low-field limit. Quantum effects [1], unusual topological band structures [2], and inhomogeneities that lead to wandering current paths [3, 4] can induce a crossover from quadratic to linear magnetoresistance with increasing magnetic field. Here we explore a series of metallic charge- and spin-density-wave systems that exhibit extremely large positive linear magnetoresistance. By contrast to other linear MR mechanisms, this effect remains robust down to miniscule magnetic fields of tens of Oersted at low temperature. We frame an explanation of this phenomenon in a semi-classical narrative for a broad category of materials with partially-gapped Fermi surfaces due to density waves.**

The evolution of the electronic transport under the influence of a magnetic field can be deeply revealing of the fundamental properties of a material. Classically, the transverse magnetoresistance $\rho_{xx}(H_z)$ evolves quadratically with magnetic field $H$, saturating at high fields if the hole and electron densities are not compensated [5-7].

The asymptotic behavior in the low field limit of $\omega_c \tau \ll 2\pi$, where $\omega_c$ is the cyclotron frequency and $\tau$ is the relaxation time, arises from the Onsager reciprocity relation, which requires $\sigma_{ij}(\mathbf{H}) = \sigma_{ji}(-\mathbf{H})$. The leading term in the MR at low field is thus $\Delta\rho = \rho(H) - \rho(H=0) \sim H^2$ [5, 6].

Other functional forms have been observed, most notably the symmetric linear form of $\Delta\rho \sim |H|$, which is non-analytic at $H=0$ but is allowed in the high-field limit, $\omega_c \tau \gg 2\pi$. There are several well-established electronic and geometric mechanisms for generating this behavior in the high-field limit [1-10], which can be separated into two classes. The first is due to special features at the Fermi surface, such as sufficiently low electron density to permit condensation of the electrons into the lowest spin-split Landau level [1], linear dispersion from a Dirac cone with infinitesimally small carrier mass [2], Umklapp scattering [9], and magnetic breakdown [7, 10]. The second class is predominantly geometric in nature, such as an average over a combination of open and closed electron trajectories in polycrystals [1, 5], guiding center diffusion in weak disorder [4], or strong mobility fluctuations due to irregular current paths from a strongly inhomogeneous medium or conductive grain boundaries [3]. Neither class addresses linear MR in the low-field regime where $\omega_c \tau \ll 2\pi$, the focus of our experiments.

While magnetoresistance in metallic paramagnets and ferromagnets is typically small and negative, as the magnetic field quenches spin fluctuations and reduces spin scattering of itinerant carriers, large positive linear MR has been reported over the years in systems with charge-density-wave (CDW) order. This includes $NbSe_2$ [10], $Nb_3Te_4$ [11], $(PO_2)_4(WO_3)_{2m}$ [12], and $R Te_3$ ($R$=Tb & Ho) [13]. The linear or sublinear MR in the low field limit has been attributed to a wide range of mechanisms, from magnetic breakdown [10] to scattering of CDW fluctuations at Fermi surface hot spots [13]. Here we argue that there exists a universal mechanism to create linear MR in the low field/low temperature limit. The salient features are a consequence of itinerant carriers turning sharp corners of the Fermi surface. This mechanism is greatly enhanced in both charge and spin-density-wave (SDW) systems, as the formation of correlated electronic states opens a gap at the Fermi surface, removing sheets of open electron paths while keeping only small electron/hole pockets/ellipsoids with small orbits of sharp curvature. The change in the Fermi surface topology provides a *generic* approach to creating linear MR in the low field limit in correlated electron systems.

We present detailed data on the density-wave systems GdSi, Cr, and SrAl$_4$, in addition to comparisons to NbSe$_3$ and (PO$_2$)$_4$(WO$_3$)$_{2m}$ derived from the literature. GdSi and Cr have SDWs driven by Fermi surface nesting with Néel transitions at $T_N$ = 54.5 K and 311.5 K, respectively [14-18]. SrAl$_4$ develops a CDW state below $T_{CDW}$ = 245 K [19], with no confounding contributions from spin order. All of the SDW and CDW ordering wave vectors are incommensurate: **Q** ~ (0, 0.483, 0.092) in GdSi [14-16], **Q** ~ (0.952, 0, 0) in Cr [18], and **Q** ~ (0, 0, 0.11) in SrAl$_4$ [20]. GdSi turns into a field-induced ferromagnet at $H_c$ = 20.7 T [15], while the SDW in Cr extends to beyond 16 T [17]. We measure $\rho_{xx}(H_z, T)$ over a wide magnetic field range (beyond $H_c$ for GdSi) and temperatures from 0.35 to 350 K. We also characterize germane Fermi surface features of GdSi via de Haas-van Alphen (dHvA) measurements at the National High Magnetic Field Laboratory.

All the listed materials exhibit a resistivity anomaly at the density wave (DW) ordering temperature for both zero and high fields (Fig. 1). In addition to the DW anomaly, these systems also show a finite-field resistivity anomaly $\rho_{xx}(T, H_z$=constant) in the zero temperature limit, which grows with decreasing $T$ and extrapolates to its greatest value at $T$=0. The onset temperature for the resistivity anomaly is marked by an arrow in Fig. 1. Similar high-field, low-temperature $\rho(T, H$=constant) anomalies have been reported in semimetallic Bi, graphite, and WTe$_2$ [21-22], where they have been ascribed to metal-insulator transitions [21]. The finite field, low temperature anomalies we consider here exist in highly-conductive CDW and SDW systems with only partially-gapped Fermi surfaces, and in many cases appear to be closely related to the dominant linear (or even sublinear) MR.

We use the pure SDW system GdSi and the pure CDW system SrAl$_4$ as our main examples. In both materials, a strong magnetoresistance with a positive, (sub)linear field dependence was observed in the low-temperature limit (Fig. 2). The magnitude of the effect is large: a MR ratio $\Delta\rho_{ii}(H)/\rho_{ii}(0)$ = 3.5 to 8 ($i$=$a, b, c$) at $H$ = 20 T for GdSi and ~75 at $H$ = 14 T for SrAl$_4$, with $\Delta\rho(H)/\rho(0)$ still ~ 1-2% at a few tens of mT. The (sub)linear behavior spans more than three decades of field range, and is clearly present even for $\omega_c\tau \ll 2\pi$.

For both GdSi and SrAl$_4$, the linear MR in the low-temperature limit evolves into a parabolic field-dependence at more elevated temperatures (Fig. 3), similar to $R$Te$_3$ in the literature [13]. Although magnetoresistance is often fit to a power-law with a temperature-dependent exponent [10, 13], our attempts to perform a scaling analysis such as the Kohler plot to collapse all $\rho(T, H)$ data onto a universal curve were not successful. Instead, there likely exist contributions from multiple electron- and hole-like conduction bands at the gapped Fermi surface, with potentially different functional forms under field.

We adopt a phenomenological approach to disentangle linear and quadratic components, fitting the measured MR as $\rho(H, T) = \rho(T, H=0) + A(T)|H| + B(T)H^2$ as a function of field at various temperatures. We note that this form is distinct from a simple crossover between linear and quadratic dependencies, and that this procedure is only valid to a magnitude of $\Delta\rho$ that is comparable to $\rho(H=0)$. As plotted in Figs. 3b and 3d for GdSi and SrAl$_4$, respectively, contributions from $A(T)$ and $B(T)$ to the magnetoresistance are very different, although they are both insensitive to the range of fitting for 0.5 T < $H$ < 4 T in GdSi, and for 0.1 T < $H$ < 0.7 T in SrAl$_4$. In both systems, the linear component $A(T)H$ dominates over the quadratic term $B(T)H^2$ at low temperature, but $A(T)H$ drops dramatically with increasing values of the zero-field resistivity, $\rho(T, H=0)$. We plot $A(T)H$ and $B(T)H^2$ against $\rho(T, H=0)$ instead of $T$ itself in Figs. 3b, and 3d to highlight the monotonic behavior of $A(\rho(H=0))$ ($A(T)$ saturates at low temperature). By comparison, $B(T)$ does not change appreciably over a wide range of temperatures. The strong linear behavior approaching the zero-temperature limit is consistent with observations regarding Nb$_3$Te$_4$ and (PO$_2$)$_4$(WO$_3$)$_{2m}$ [11-12].

The rapid diminution of $A(T)$ at elevated temperature argues against phonon-[9] or excitation-based [13] scattering mechanisms as sources for the linear MR. It is also important to note that the low-temperature resistivity anomaly in $\rho(T, H=$constant$)$ becomes more prominent in clean systems with higher residual resistivity ratio (RRR) values. In NbSe$_3$, for example, the resistivity anomaly was only observed in samples with RRR values of 284 and 87, but not of 31 [23].

The similarity in behavior between CDW and SDW systems (Figs. 1-3) suggests that both the low-temperature anomaly and the low-field MR, $\Delta\rho \sim |H|$, may arise from a generic mechanism that is intrinsic to general features of the Fermi surface. The gapped Fermi surface of SrAl$_4$ has several ellipsoids whose sizes are as small as 0.5% of the Brillouin zone volume [19]. We have performed density functional theory calculations for GdSi and, as illustrated in Fig. 4a, we find that its Fermi surface is marked by closed, bowtie or pinwheel-like volumes. These small volumes and their associated tight electron orbits contrast with the paramagnetic phase which has extended Fermi surfaces with nesting characteristics (Fig. 4b) [14, 16].

To validate our band structure calculations for GdSi, we probe the Fermi surface via dHvA oscillations (Fig. 4). The lowest frequency measured is 71.5 T for $H$ along the $c$-axis, which was also detected through a significant part of the $b$-$c$ plane (Fig. 4f). The slowly varying form of $F(\theta)$ suggests an ellipsoidal Fermi surface that should present a small orbital for $H$ along the $b$-axis. The 71.5 T frequency represents an orbital of lateral size 0.05-0.1 Å$^{-1}$, and enclosing 0.53% of the cross-sectional area of the first Brillouin zone in the $a$-$b$ plane.

Both GdSi and SrAl$_4$ have similarly high electron densities $n$ in the gapped phase, with Hall coefficients $R_H$ of order several 10$^{-10}$ m$^3$/C (Refs. [14, 19]). With $\rho(T=1.7K, H=0)$ of GdSi and SrAl$_4$, of 2.2, and 0.31 μΩ–cm, respectively, and carrier masses all around unity, the cyclotron frequency $\omega_c$ is typically small. For GdSi with seven conduction electrons per formula unit (three Gd 5$d$-6$s$ and four Si 3$s$-3$p$ electrons), we estimate $\omega_c \tau = H/\rho nec = 4\times10^{-3}$ at $H = 2$ T for our sample. The criteria for quantum linear magnetoresistance [2] is thus not satisfied under our measurement conditions. Additionally, itinerant carriers would not have time to circle a full Fermi surface orbital. Instead they would only follow a small arc before being scattered away by either phonons or disorder to an incoherent state on the Fermi surface.

In general, a small arc would induce a very small directional change to the propagation of the carrier. However, the local curvature of the orbit could become large enough that within $\tau$ a carrier could round a sharp corner to change its direction by a significant fraction of $2\pi$ before being scattered away. The turning process at sharp corners would dominate the magnetoresistive contributions from all other carriers at flat parts of Fermi surface because the curvature of a sharp corner could define a large but local angular frequency $\omega$. Despite the corner representing only a portion of a full revolution, the most significant contribution to the MR would come from the effectively large $\omega_c \tau$. Carrier movement on other, flatter regions of the Fermi surface only weakly contributes to the MR with a conventional quadratic dependence.

Although the importance of local curvature has been recognized in the literature, most often it still leads to predictions of quadratic MR because of assumptions of finite-sized rounding of the Fermi surface feature [24-25], perfectly suited to explain the large

MR in Bi, graphite, and WTe$_2$ with their respective small Fermi surface pockets [21-22]. However, Pippard pointed out that there exists linear MR for a square Fermi surface with an infinitely sharp corner (pages 35-39 of Ref. [7]). Intuitively, this linear dependence can be understood by noting that in the reciprocal space, the global cyclotron frequency, $\omega_c$, linearly increases with $H$, while the orbital size remains unchanged. Thus, the number of electrons moving along the orbital of the Fermi surface that can access the sharp corner within the mean scattering time before colliding, $\tau$, increases linearly with field, and so does the magnetoresistance. The MR behaves linearly as long as the mean free pathlength in reciprocal space, $l_k$, is much larger than the radius of the sharp corner, $r_k$, so that rounding is not relevant.

The field limit below which $\Delta\rho/\rho$ is no longer observable serves to estimate the curvature of the sharpest corner on the Fermi surface. The Onsager-Lifshitz relation connects orbit radii in reciprocal space, $r_k$, and real space, $r$, by $r_k = \frac{eH}{c\hbar}r$. When the carriers' mean free path, $l$, becomes comparable to $r$ at the corner with decreasing $H$, the linear MR disappears. With disorder, the sharp turning points in the band structure could be broadened on the scale of the mean free path: $r_k \sim 1/l$. Using these two expressions for $r_k$, and taking $l\sim1000$Å and $H$=10-100Oe, we estimate $r_k \sim 10^{-4}$ Å$^{-1}$ in the SrAl$_4$ and GdSi systems.

The Pippard approach does not address how sharp corners in the band structure survive in real materials, especially as different branches of the original Fermi surface hybridize via the CDW. An alternative way to create sharp band structure features in a CDW or SDW may arise from avoided crossings of nested pockets. The momentum required to scatter from one branch of the Fermi surface to an (approximately) orthogonal branch is extracted by Umklapp scattering from the CDW or SDW order. Hence a (possibly disorder-assisted) tunneling process could play the role required by Pippard's mechanism to act as a source/sink of particles on the two branches. This would be a nonlinear process in terms of the ratio of the impurity potential to the gap – i.e. essentially using strong coupling impurities to reform the (hidden) Fermi surface of the parent metal, and to scatter between the intersecting branches. The idea is similar to the "Umklapp" mechanism, except that it would not require thermal excitation of particle-hole pairs.

It follows that the competition between coherent turning of sharp corners on the Fermi surface and incoherent collisions with phonons is the origin of the high-field resistivity anomaly in $\rho(T, H$=constant) exhibited in Fig. 1. $\tau$ decreases by collisions with either disorder or phonons. With a decreasing mean free path with increasing temperature, fewer electrons will be able to circle the sharp corner before collisions; hence the magnetoresistance initially decreases. The MR eventually begins to increase again due to enhanced phonon scattering at elevated temperature. With a disorder-shortened $l_k$ closer to $r_k$ of the turning corner, the power-law exponent of the field dependence would be expected to change from one to two.

This mechanism for large, positive, and linear MR is well suited for explaining the behavior of a broad family of density-wave materials, and potentially other metallic incommensurate antiferromagnets in which large patches of Fermi surface are gapped out by the formation of long-range order [26]. By removing sheet-like Fermi surfaces for open electron paths upon ordering, closed electron or hole pockets represent a generic topology for the remaining density of states. Although only representing a small portion of the Fermi surface, they can manifestly dominate the response to an applied field. In fact, as seen for SrAl$_4$, its (sub)linear MR at $T = 1.7$ K is of order 7,500% by $H = 14$ T, effectively matching the total electrical resistivity at room temperature.

**Methods:**

Aligned single crystals of GdSi, Cr, and SrAl$_4$ were prepared in bar shapes of 1-3 mm length for magnetotransport measurements. Electrical leads were attached by silver epoxy in a four-lead geometry. High-field MR was measured using LakeShore LS372 resistance bridges and a LS3708 preamplifier in either an 18-Tesla superconducting magnet or a 35-Tesla DC magnet (Cell 12) at the National High Magnetic Field Laboratory (NHMFL), Tallahassee, FL, using Helium-3 cryostat inserts to reach 0.34 K. The lower-field data were taken with similar electronics in a 14 T Physical Property Measurement System (PPMS, Quantum Design, Inc.) for 1.7 K ≤ T ≤ 350 K. All low-field data were taken following protocols to remove potential trapped flux in the superconducting magnet. Nevertheless, we estimate that the zero field value is only precise to 20 Oe.

de-Haas van-Alphen (dHvA) quantum oscillations were measured using cantilever torque magnetometry with Andeen-Hagerling AH2700A capacitance bridges in an 18-Tesla superconducting magnet down to $T = 0.34$ K at the NHMFL. A monotonic polynomial background was first removed, and the oscillating spectra were Fourier transformed to detect major frequencies. The data was further fit to the Lifshitz-Kosevich formula, as the oscillating amplitude from torque magnetometry depends on temperature and field as $\sim TH^{-1/2}\exp(-\alpha T_D/H)/\sinh(-\alpha T/H)$, where $T_D$ is the Dingle temperature, and coefficient $\alpha=14.69(m/m_e)$ Tesla/Kelvin, with the carrier mass $m$ normalized by the free electron mass $m_e$. This provides the fitting functional form in Fig. 4 for both carrier mass and Dingle temperature.

Density functional theory calculation employed a 1×2×1 super-cell with a halved first Brillouin zone of non-magnetic GdSi. This procedure approximates the real **Q**~(0, 0.493, 0.092) with a commensurate (0, ½, 0) antiferromagnetic order [14].

**Acknowledgments**

We are grateful to N. Woo and J. Wang for help with the data collection, and to H. Chen for stimulating discussions. Y.F. gratefully acknowledge the support from Okinawa Institute of Science and Technology Graduate University with subsidy funding from the Cabinet Office, Government of Japan. The work at Caltech was supported by National Science Foundation Grant No. DMR-1606858. Work performed at the National High Magnetic Field Laboratory was supported by National Science Foundation Cooperative Agreement No. DMR-1157490 and the State of Florida. J.-Q. Y. was supported by the US Department of Energy, Office of Science, Basic Energy Sciences, Division of Materials Sciences and Engineering. Y. Ō. acknowledges Japan Society for the Promotion of Science KAKENHI Grant Nos. JP18H043298, JP17K05547, and JP16K05453. B.M. acknowledges support from the National Science Foundation through its employee IR/D program.

**Author contributions**

Y.F. and T.F.R. conceived of the research. J.-Q.Y., T.N., M.H. and Y.Ō. provided single crystal samples. Y.F., Y.W., D.M.S., R.K., A.S. performed measurements. B.M. performed band structure calculation. Y.F., Y.W., and P.B.L. developed the theoretical framework. Y.F., Y.W., D.M.S., and T.F.R. analyzed the data and prepared the manuscript. All authors commented on the manuscript.

**Additional information**

Correspondence and requests for materials should be addressed to Y.F. <yejun@oist.jp>

or T.F.R. <tfr@caltech.edu>.

**Competing financial interests**

The authors declare no competing financial or non-financial interests.

**Figure captions:**

**Fig. 1 Finite-field magnetoresistance anomaly in CDW/SDWs.** Transverse magnetoresistance $\rho_{xx}(T, H_z)$ at zero (blue) and high (red) fields are contrasted for (**a**, **b**) GdSi (**H** || *a* and *b*, respectively), (**c**) SrAl$_4$, (**d**) Cr, (**e**) (PO$_2$)$_4$(WO$_3$)$_{2m}$ (adapted from Ref. [12]), and (**f**) NbSe$_3$ (adapted from Ref. [23]). The CDW and SDW transitions manifest as the resistance anomalies at higher temperature. Arrows point to where the in-field low-temperature resistance anomaly starts to grow with decreasing temperature. SrAl$_4$ has an extra structural transition at $T \sim 80$ K as indicated by the temperature hysteresis [19]. For our samples, the ratio of $\rho(300K)/\rho(2K)=25$ for GdSi with $I//b$, 58 for GdSi with $I//c$, 77 for SrAl$_4$, and $\rho(350K)/\rho(2K)=92.4$ for Cr. This resistance anomaly at finite fields and zero temperature was also observed in semi-metals Bi, graphite [21], and WTe$_2$ [22].

**Fig. 2. Linear transverse magnetoresistance at low field.** Positive linear magnetoresistance over a field range spanning more than three decades, with (**a**, **d**) $\Delta\rho/\rho \sim 800\%$ in GdSi and $\sim 7,500\%$ in SrAl$_4$; (**b**, **e**) $\Delta\rho/\rho \sim 100\%$; and (**c**, **f**) $\Delta\rho/\rho \sim 1\%$. Black solid and dashed lines are even-parity linear ($\sim |H|$) and quadratic fits to the data respectively. Shubnikov-de Haas oscillations are noticeable at high $H$ in both systems.

**Fig. 3. Temperature evolution of linear and quadratic components.** (**a**, **c**) Representative curves of $\Delta\rho(H, T)$ are plotted together for varying temperatures. Fitting each curve with a sum of linear and quadratic terms, $\Delta\rho(H, T) = A(T)|H| + B(T)H^2$, gives the coefficients $A(T)$ and $B(T)$. (**b**, **d**) These coefficients are compared for their contributions (linear with $A(T)H$ and quadratic with $B(T)H^2$) to the magnetoresistance under a field of 2 T for GdSi and 0.5 T for SrAl$_4$, respectively.

**Fig. 4. Closed Fermi surface in the SDW state with small orbitals.** (**a**) Three-dimensional schematics of DFT-calculated Fermi surface in the spin-density-wave

phase of GdSi ($T<T_{SDW}$, Methods). All forms are closed with no open surfaces. (**b**) By comparison, the Fermi surface in the paramagnetic phase of GdSi possesses multiple sheets in reciprocal space for extended carrier paths. Figure is adapted from Refs. [14, 16]. (**c**) Quantum oscillations as a function of inverse magnetic field, expressed in measured capacitance. The fitting (solid line) reveals $F$ = 1085 T, 508 T, and 432 T, with respective Dingle temperatures $T_D$ = 2.8, 4.7, and 2.8 K. No frequency higher than 1000 T was observed along either the *c* or *a* axes. (**d**) Temperature dependence of major quantum oscillations along various crystalline axes, revealing carrier masses from 0.75 to 1.34 $m_e$. (**e**) Raw data of the detected quantum oscillation frequency of $F$ = 76 T, and (**f**) its evolution in the ***b-c*** plane with the lowest $F_0$ = 71.5 T along the *c*-axis. The short-dashed line in a functional form of $F_0/\cos(\theta)$ indicates the expectation from a tubular shaped Fermi surface, while the solid curve (and long-dashed line extension) is a fit of data to the expected form of an ellipsoidal Fermi surface.

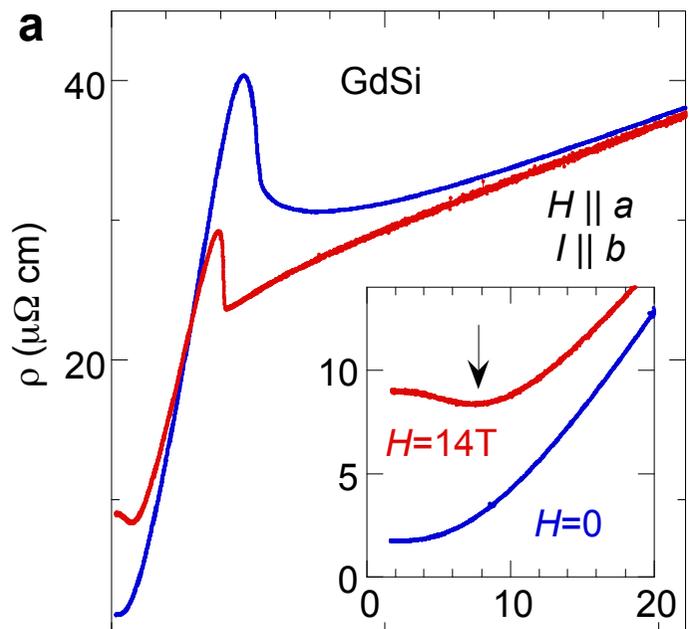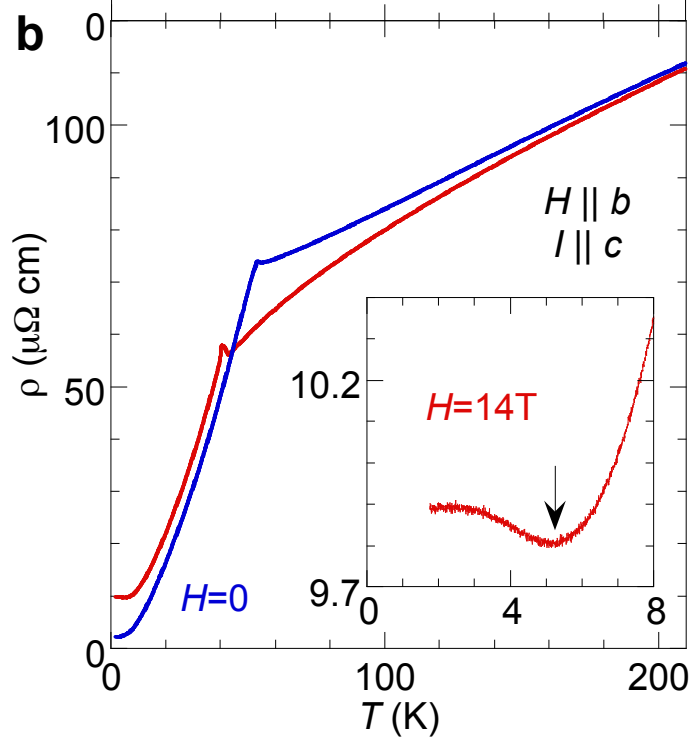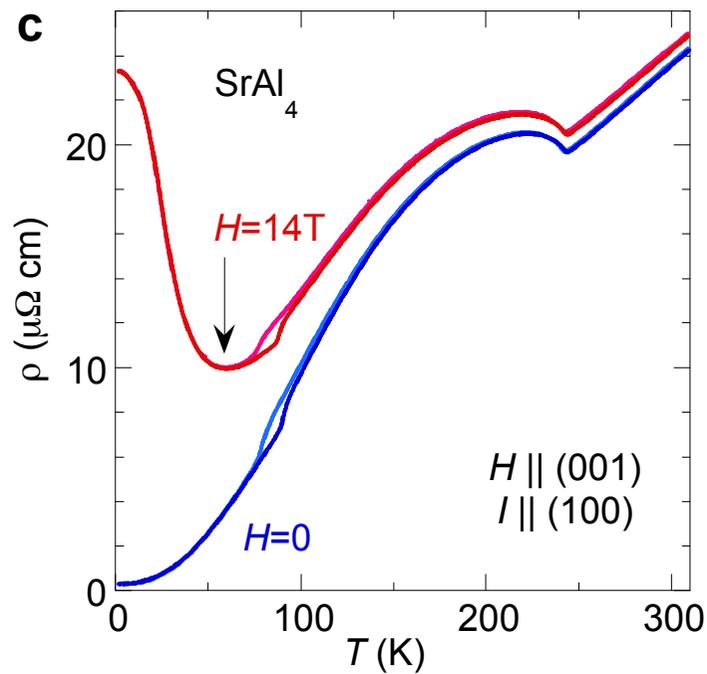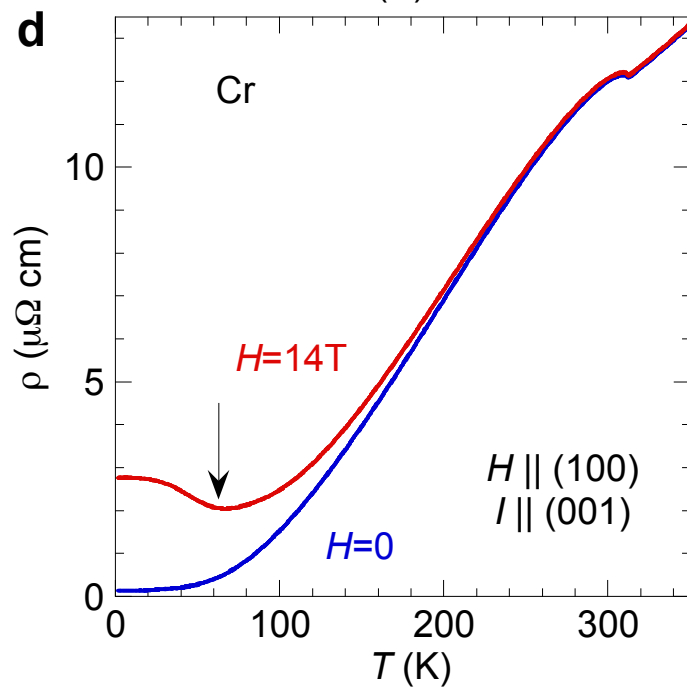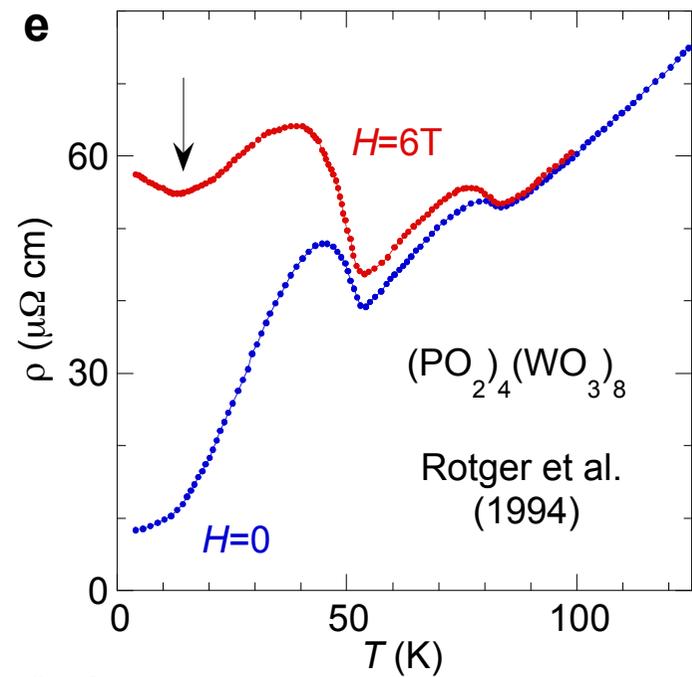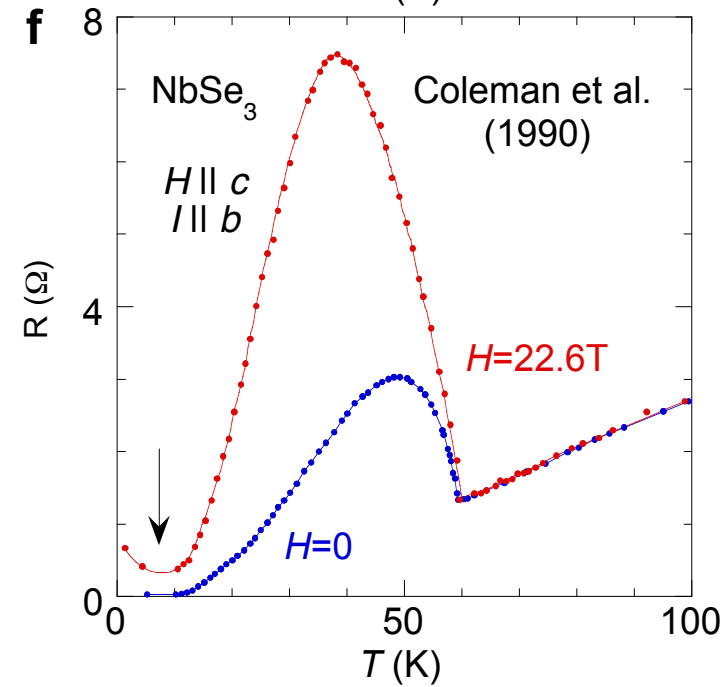

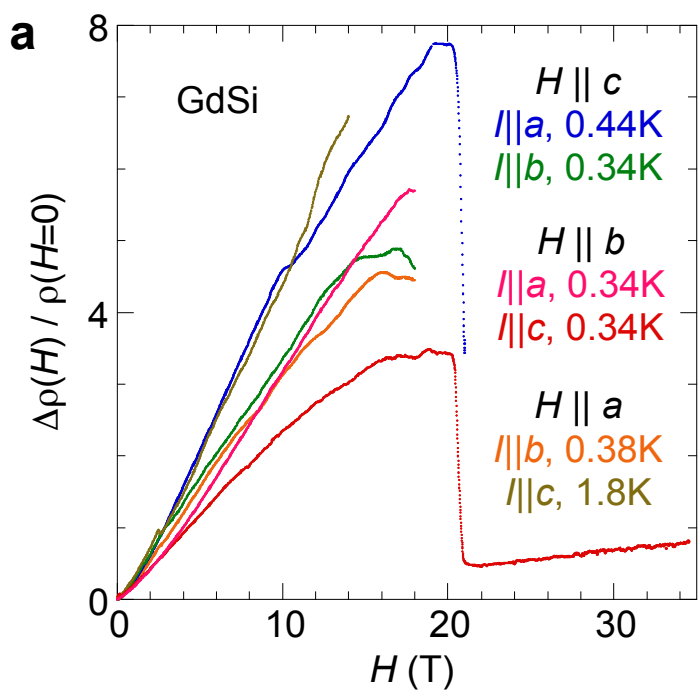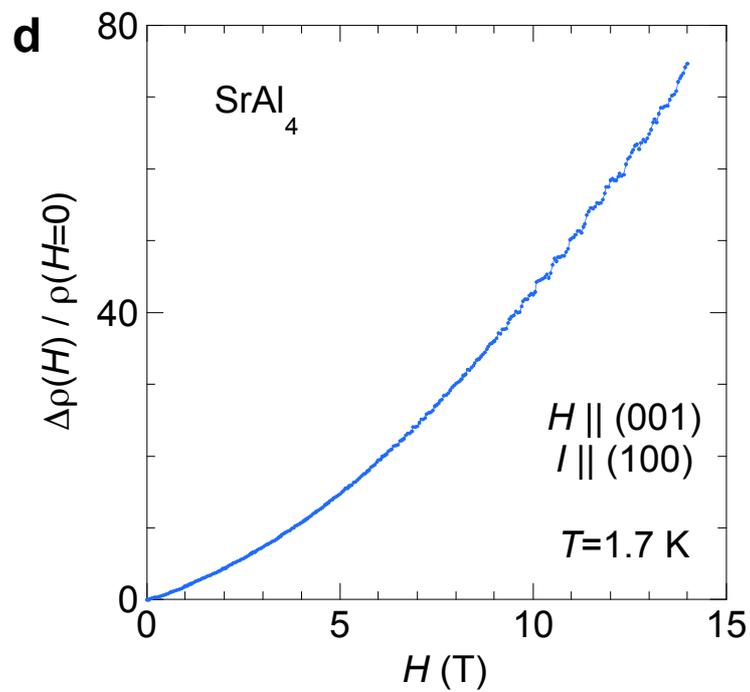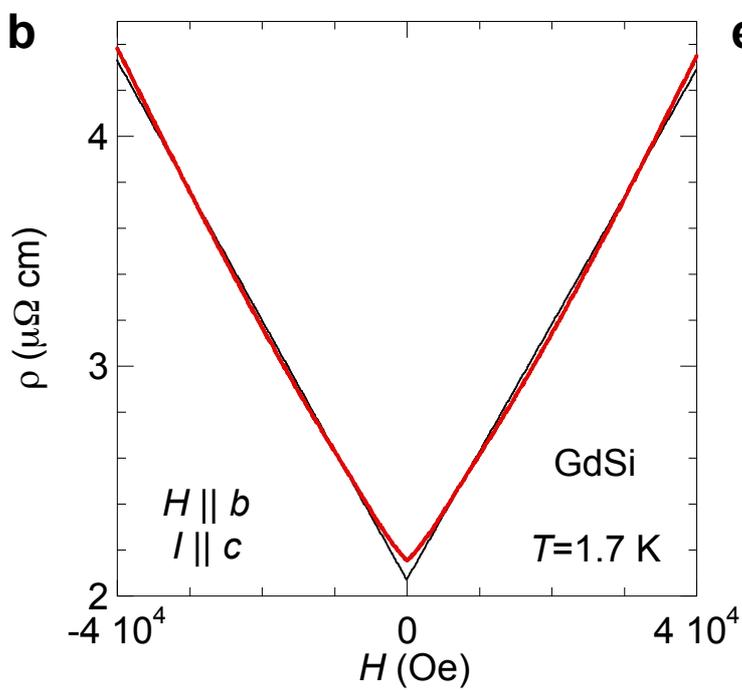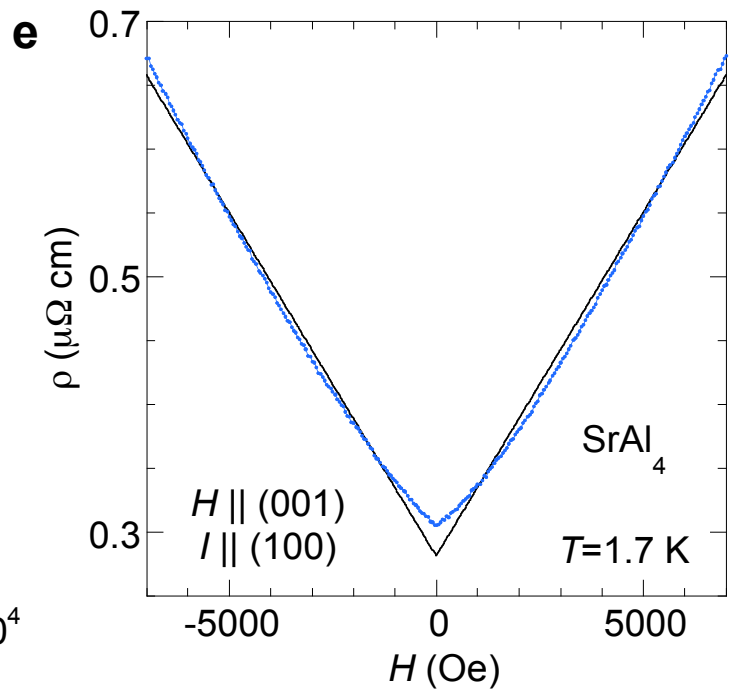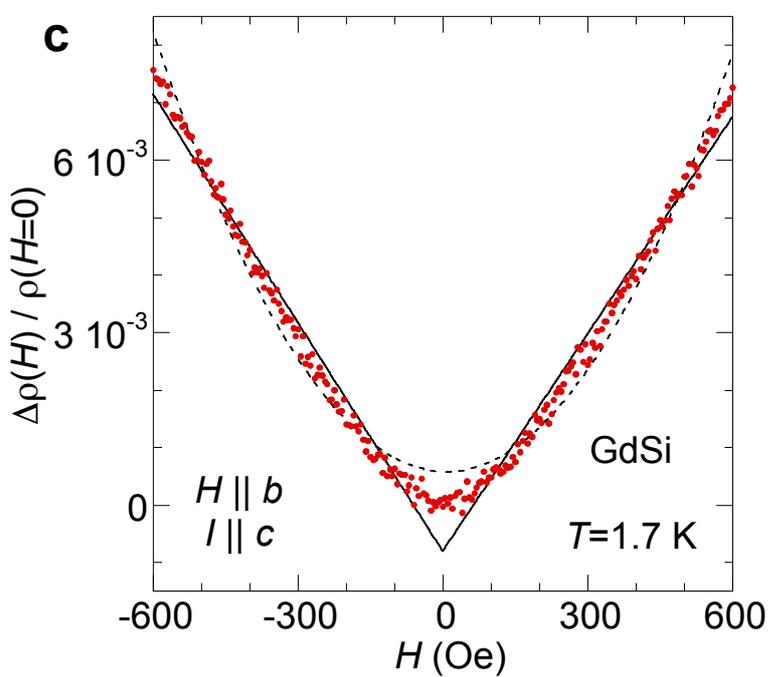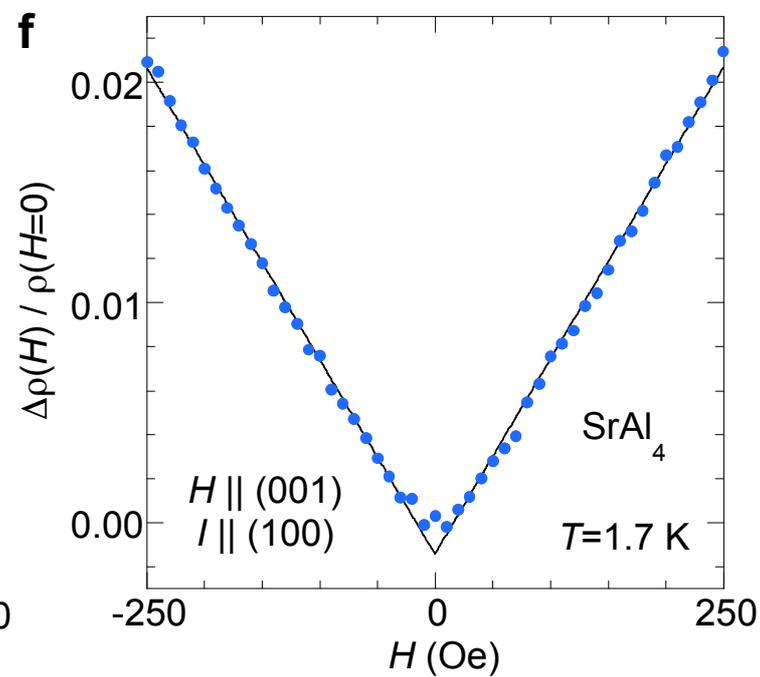

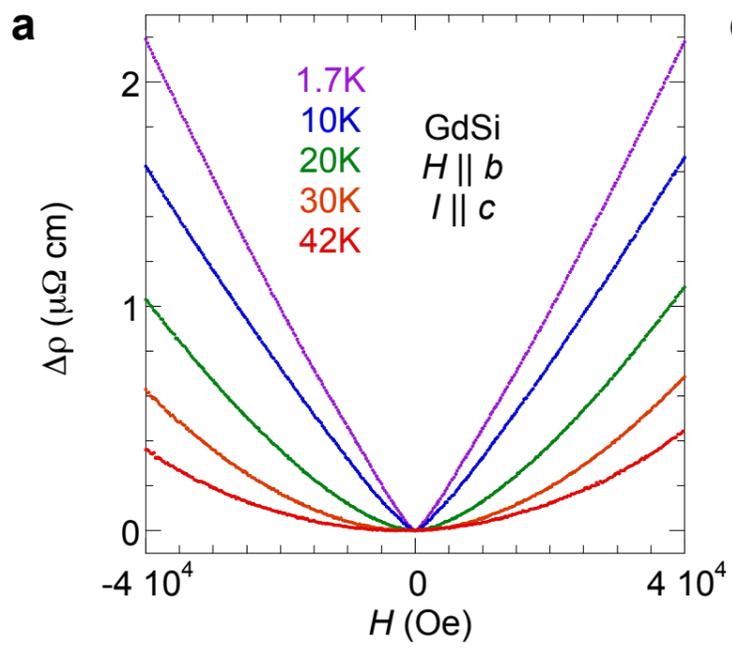
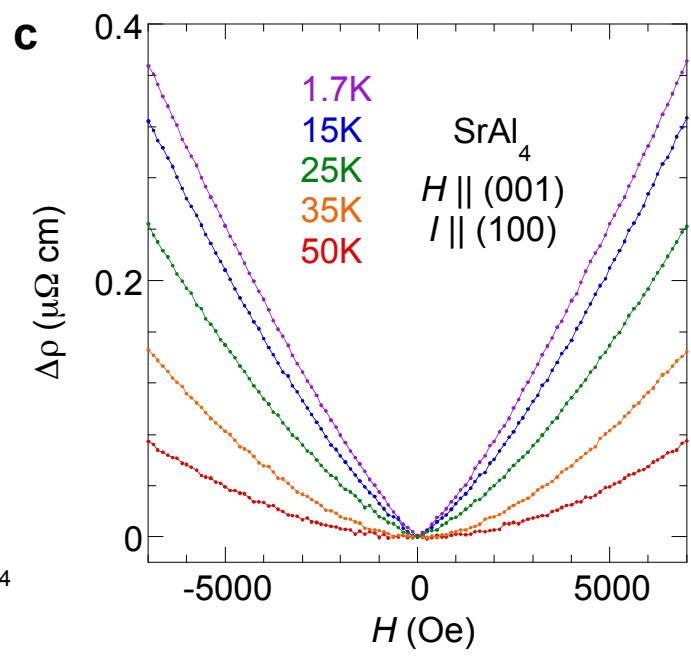
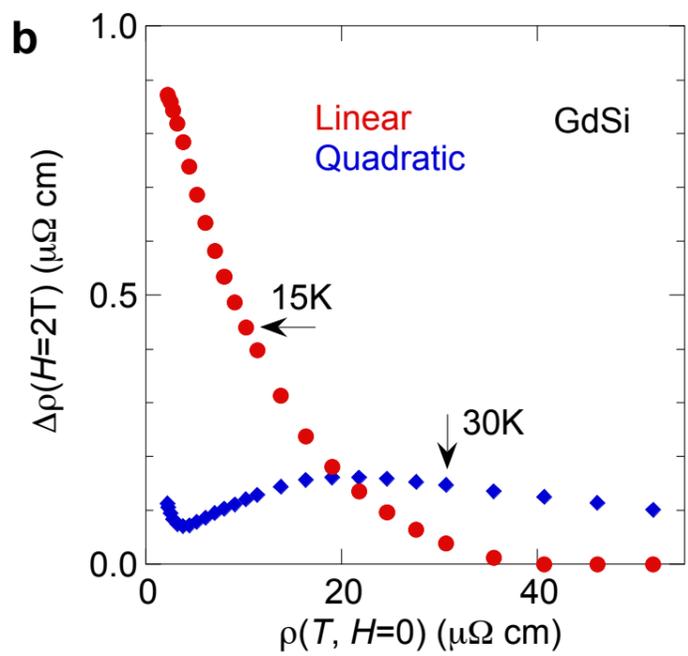
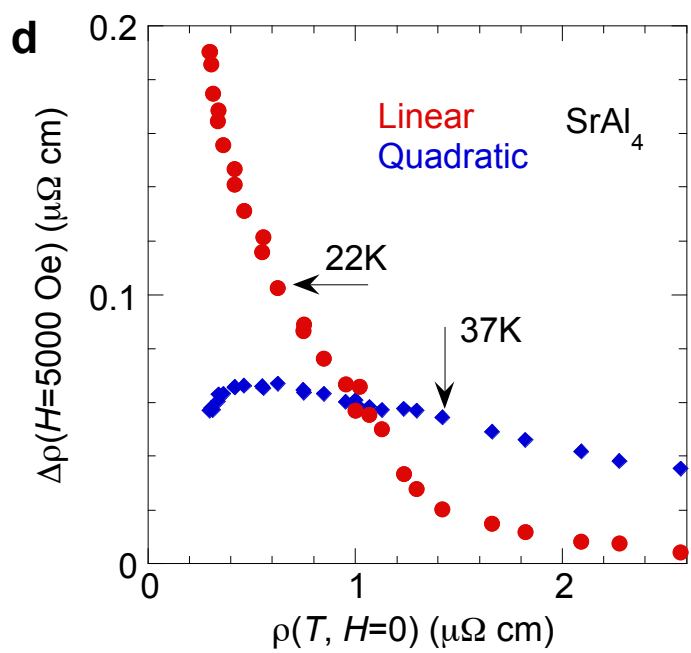

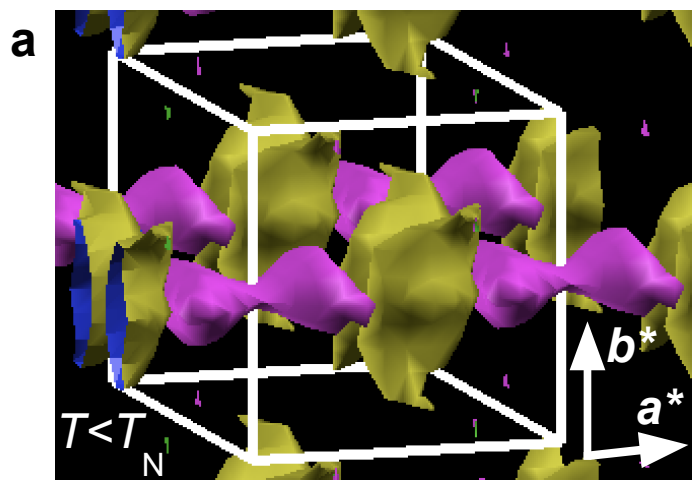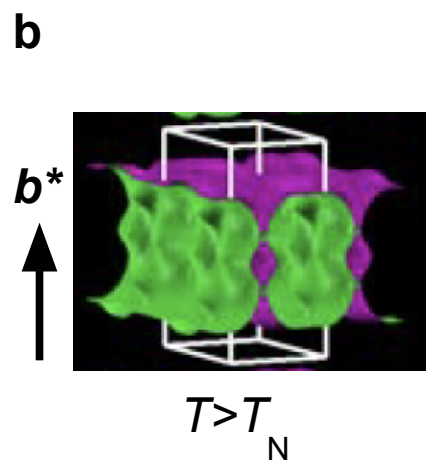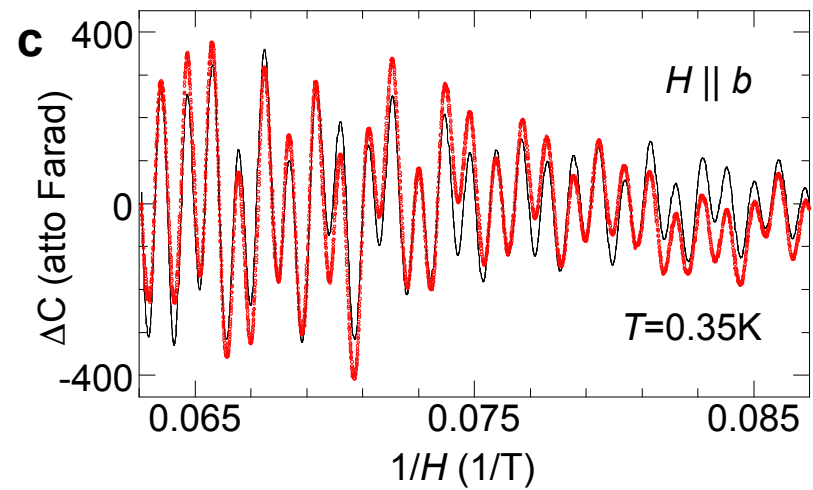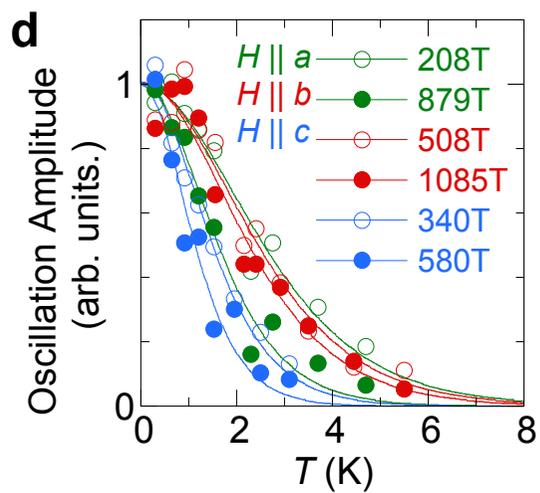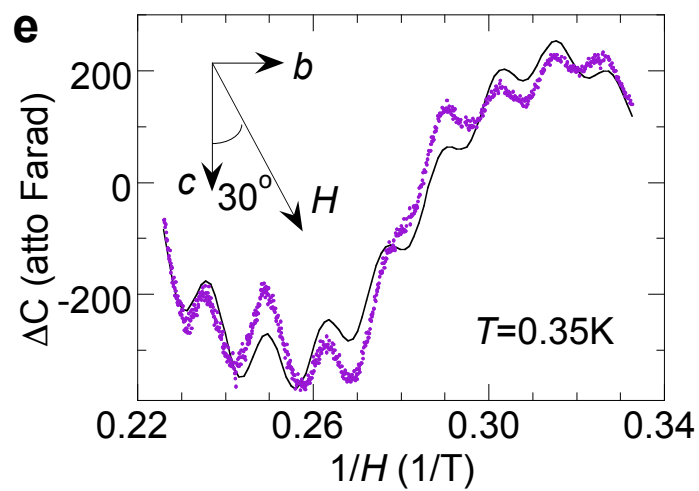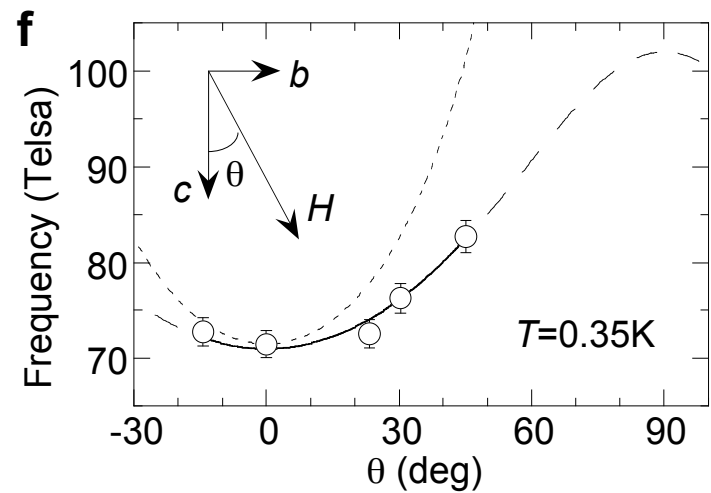